\documentclass[prx,twocolumn,superscriptaddress,longbibliography,nofootinbib]{revtex4-2}

\usepackage[utf8]{inputenc}
\usepackage{graphicx}

\usepackage[dvipsnames]{xcolor}
\definecolor{dark-blue}{rgb}{0,0.2,0.6}

\usepackage[colorlinks,%
            pdfusetitle,%
            urlcolor=dark-blue,%
            citecolor=dark-blue,%
            linkcolor=dark-blue]{hyperref}

\usepackage{etoolbox}
\makeatletter
\pretocmd{\NAT@open}{\begingroup\color{\@citecolor}}{}{}
\apptocmd{\NAT@close}{\endgroup}{}{}
\makeatother

\usepackage{stix}
\DeclareMathAlphabet{\mathcal}{OMS}{cmsy}{m}{n}
\DeclareSymbolFont{CMAlt}{OMX}{cmex}{m}{n}
\DeclareMathSymbol{\sumop}{\mathop}{CMAlt}{"50}
\DeclareMathSymbol{\intop}{\mathop}{CMAlt}{"52}

\usepackage{amsmath}
\usepackage{upgreek}
\newcommand{\yb}{\ensuremath{{^\text{174}\text{Yb}}}}
\newcommand{\ybo}{\ensuremath{{^\text{171}\text{Yb}}}}

\newcommand{\tP}[1]{\ensuremath{{^3\text{P}_{#1}}}}
\newcommand{\tD}[1]{\ensuremath{{^3\text{D}_{#1}}}}

\newcommand{\tS}[1]{\ensuremath{{^3\text{S}_{#1}}}}

\newcommand{\sS}[1]{\ensuremath{{^1\text{S}_{#1}}}}

\newcommand{\sP}[1]{\ensuremath{{^1\text{P}_{#1}}}}

\newcommand{\mathunit}[1]{\ensuremath{\,#1}}
\newcommand{\asciimathunit}[1]{\ensuremath{\,\text{#1}}}

\newcommand{\nm}{\asciimathunit{nm}}
\newcommand{\um}{\mathunit{\upmu\text{m}}}
\newcommand{\mm}{\asciimathunit{mm}}

\newcommand{\mHz}{\asciimathunit{mHz}}
\newcommand{\Hz}{\asciimathunit{Hz}}
\newcommand{\kHz}{\asciimathunit{kHz}}
\newcommand{\MHz}{\asciimathunit{MHz}}
\newcommand{\GHz}{\asciimathunit{GHz}}
\newcommand{\THz}{\asciimathunit{THz}}

\newcommand{\mW}{\asciimathunit{mW}}

\newcommand{\Gauss}{\asciimathunit{G}}
\newcommand{\mG}{\asciimathunit{mG}}

\newcommand{\uK}{\mathunit{\upmu\text{K}}}

\newcommand{\s}{\asciimathunit{s}}
\newcommand{\ms}{\asciimathunit{ms}}
\newcommand{\us}{\mathunit{\upmu\text{s}}}
\newcommand{\ns}{\asciimathunit{ns}}

\newcommand{\Erec}{E_\text{rec}}

\newcommand{\subfigref}[2]{\hyperref[fig:#1]{\ref*{fig:#1}(#2)}}

\usepackage{xr}

\begin{document}


\title{Determining the $\tP{0}$ excited-state tune-out wavelength of $\yb$ in a triple-magic lattice}

\author{Tim~O.~H\"ohn}
\author{Ren\'e~A.~Villela}
\author{Er~Zu}
\author{Leonardo~Bezzo}
\author{Ronen~M.~Kroeze}

\affiliation{Fakult{\"a}t f{\"u}r Physik, Ludwig-Maximilians-Universit{\"a}t, 80799 M{\"u}nchen, Germany}
\affiliation{Munich Center for Quantum Science and Technology (MCQST), 80799 M{\"u}nchen, Germany}

\author{Monika~Aidelsburger}
\email{Monika.Aidelsburger@physik.uni-muenchen.de}

\affiliation{Fakult{\"a}t f{\"u}r Physik, Ludwig-Maximilians-Universit{\"a}t, 80799 M{\"u}nchen, Germany}
\affiliation{Munich Center for Quantum Science and Technology (MCQST), 80799 M{\"u}nchen, Germany}
\affiliation{Max-Planck-Institut f{\"u}r Quantenoptik, 85748 Garching, Germany}


\date{\today}

\begin{abstract}
Precise state-dependent control of optical potentials is of great importance for various applications utilizing cold neutral atoms. 
In particular, tune-out wavelengths for the clock state pair in alkaline-earth(-like) atoms provide maximally state-selective trap conditions that hold promise for the realization of novel approaches in quantum computation and simulation. 
While several ground-state tune-out wavelengths have been determined, similar experimental studies for metastable excited states are challenged by inelastic collisions and Raman losses, so far prohibiting precise measurements of excited-state tune-out conditions.
In this work we report on the measurement of a tune-out wavelength for the metastable $\tP{0}$ clock state in $\yb$ at $519.920(9)\,$THz.
In order to circumvent collisional losses, we isolate individual $\tP{0}$ atoms in a clock-magic-wavelength lattice at $759\,$nm. 
To minimize the limitation imposed by Raman scattering, we further implement resolved sideband cooling on the clock transition, which allows us to reduce the lattice depth and surpass lifetimes of $5\,$s. 
The precision of the tune-out measurement is further enhanced by fluorescence imaging in a triple-magic configuration, where we implement molasses cooling on the $\tP{1}$ intercombination line and identify a magic angle of $38.5(9)^\circ$ in the clock-magic lattice.
\end{abstract}

\maketitle


\section{Introduction}
Alkaline-earth(-like) (AEL) atoms have been utilized for a plethora of remarkable applications, from advanced quantum computation protocols~\cite{lis:2023,ma:2023,huie:2023,reichardt:2024} to quantum simulation of the SU($N$) Fermi-Hubbard model~\cite{scazza:2014,pagano:2014,taie:2022,pasqualetti:2024}.
In particular the ultra-narrow transition to the $\tP{0}$ state has enabled the development of extremely accurate optical lattice clocks~\cite{ye:2008,ludlow_optical_2015,aeppli:2024}, the realization of highly entangled states for enhanced metrology~\cite{cao:2024}, as well as the realization of spin-orbit coupling~\cite{kolkowitz:2017} and artificial gauge fields in synthetic dimensions~\cite{mancini:2015,zhou:2024}.
A crucial ingredient for these results is the precise cancellation of differential Stark shifts in magic traps~\cite{barber:2008}.

However, for certain applications like site-resolved addressing~\cite{lis:2023,chen:2023,bluvstein:2024} or the simulation of mass-imbalanced particles~\cite{darkwahoppong:2022} and lattice gauge theories~\cite{surace_abinitio_2023}, state-selective traps constitute an invaluable resource.
The extreme case of a fully vanishing atom-light coupling of one state at a tune-out wavelength, while the other state retains a finite polarizability, has been employed for erasure conversion of quantum gate errors~\cite{reichardt:2024}, and it has been suggested to be used for simulations of twisted bilayer systems~\cite{gonzalez-tudela:2019,luo_spin-twisted_2021}, the realization of novel quantum gates~\cite{pagano_fast_2019}, and the separation into storage and transport traps for quantum processors~\cite{daley:2008,gonzalez-cuadra_fermionic_2023}.
Since this condition requires equal blue- and red-detuned polarizability contributions from the dominant transitions, tune-out wavelengths in alkali atoms only exist close to transitions, resulting in detrimental near-resonant scattering.
For AEL atoms, the strongly decoupled transition manifolds for the ground and metastable excited state in turn result in far-detuned tune-out wavelengths and thus provide the ability to perform high-fidelity resorting operations or local lightshift applications with minimal scattering.

\begin{figure}[ht!]
	\includegraphics{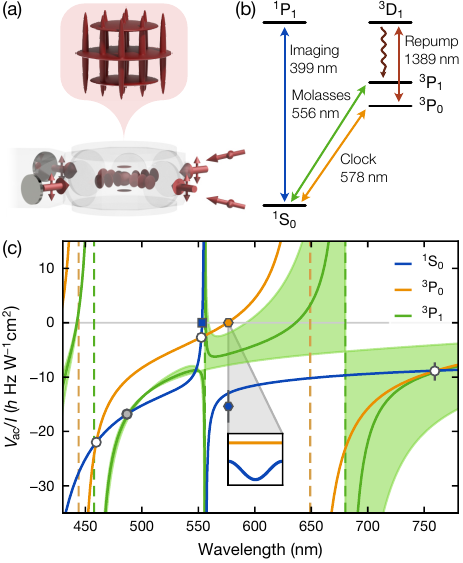}
	\caption{\label{fig:1}
		\textbf{Experimental setup, simplified level structure, and ac polarizabilities of $\yb$}.
		$\left(\text{a}\right)$ Sketch of the 3D clock-magic lattice. Double-sided arrows indicate the polarization of the individual lattice beams.
		$\left(\text{b}\right)$~Simplified level scheme including the clock ($\sS{0}\to \tP{0}$), repumping ($\tP{0}\to \tD{1}$), imaging ($\sS{0}\to \sP{1}$) and molasses cooling transition ($\sS{0}\to \tP{1}$).
		$\left(\text{c}\right)$ Light shifts of the three lowest-lying states in linearly polarized traps, using an empirical model~\cite{hoehn:2023,SM}. The green shaded area highlights the tunability of the total $\tP{1}(m_{J'}=0)$ light shift by adjusting the angle of the quantization axis. Round markers indicate the measured magic wavelengths ($\sS{0}\to \tP{0}$ white, $\sS{0}\to \tP{1}$ gray), hexagonal markers the $\tP{0}$ tune-out wavelength (orange, corresponding measured $\sS{0}$ polarizability in blue) and the square marker depicts the $\sS{0}$ tune-out wavelength.
        Inset: Schematic trapping potential at the $\tP{0}$ tune-out wavelength.}
\end{figure}

Furthermore, as precise $\textit{ab initio}$ polarizability calculations for the $\sS{0} - \tP{0}$ clock state pair are highly nontrivial, measurements of distinct points such as tune-out wavelengths are crucial to test and improve the accuracy of these models~\cite{safronova_extracting_2015}.
In this work we demonstrate the versatility of shaping the polarizabilities of various states by measuring the magic angle for the narrow $\sS{0} \to \tP{1}$ cooling transition in $\yb$, allowing for a triple-magic condition in the clock-magic lattice at $759 \nm$, and determining the tune-out wavelength for the $\tP{0}$ clock state, where its polarizability exhibits a zero crossing.

Compared to the ground-state tune-out measurements performed in AEL atoms so far, which have been based on parametric heating schemes with long-lived ground-state atoms trapped in one dimensional (1D) lattices~\cite{heinz:2020,hoehn:2023}, the detection of a tune-out wavelength for the metastable $\tP{0}$ state is significantly challenged by several factors.
First, strong inelastic collisions of nearby $\tP{0}$ atoms result in fast losses~\cite{franchi:2017,bouganne:2017}, limiting Kapitza-Dirac scattering measurements using pulsed optical lattices with a Bose-Einstein condensate (BEC)~\cite{herold_precision_2012,ratkata:2021,schmidt_precision_2016,kao:2017,catani:2009}.
Second, deep optical traps induce strong off-resonant Raman scattering~\cite{dorscher:2018,SM}, such that one has to resort to shallow optical lattices to reach the sufficiently long lifetimes necessary for parametric heating measurements, as used in this work~\cite{heinz:2020,hoehn:2023}.
To this end, we combine ground-state cooling of bosonic $\yb$ atoms on the clock transition in a 2D lattice~\cite{kroeze:2024} with a high signal-to-noise ratio detection scheme using magic-angle molasses cooling during fluorescence imaging, where the former allows for a sufficient isolation of individual atoms and suppresses tunneling, while the latter provides the resolution to detect even smallest lifetime changes.

\section{Experimental results}\label{sec:sec1}
The experiment starts by directly loading $\simeq \! 70 \times 10^3$ $\yb$ atoms from a magneto-optical trap (MOT) into a $\simeq \! 140 \uK$ deep clock-magic optical lattice at $\lambda = 759.3 \nm$.
Compared to our previous work, our setup has been upgraded to a 3D lattice, which is used for loading and imaging of the atoms~\cite{hoehn:2023}. 
The 3D geometry consists of two orthogonal, retro-reflected and vertically polarized horizontal lattices and a shallow-angle vertical lattice as illustrated in Fig.~\subfigref{1}{a}.
The 3D lattice allows us to reach total potential depths of $\simeq \! 450 \uK$ and strong confinement in all directions.
Fluorescence imaging is realized by implementing molasses cooling on the $\sS{0} \to \tP{1}$ intercombination line [Fig.~\subfigref{1}{b}]~\cite{yamamoto:2016,jenkins:2022,ma:2022,lis:2023}. Since this transition is only $183\,$kHz wide, high-fidelity imaging relies on magic trapping~\cite{ma:2022,norcia:2023,norcia:2024}.
This condition is generally not fulfilled in a clock-magic optical lattice.
The finite total electronic angular momentum of the $\tP{1}$ state, however, induces a significant tensor shift, which can be leveraged to achieve a triple-magic condition [Fig.~\subfigref{1}{c}], as has been demonstrated for $\ybo$~\cite{lis:2023}.
Here, one can utilize the dependence of the tensor shift on the relative angle $\theta$ between the trap polarization and the quantization axis.
In the case of $\yb$, the total expected resonance shift for the Zeeman substate $m_{J'}$ in the presence of a magnetic field $B$ can be expressed as
\begin{equation}\label{eq:magiccondition}
	\Delta V\!_\mathrm{ac} = - \frac{I}{2c\epsilon_0} \left(\!\Delta \alpha_\mathrm{s} + \alpha_\mathrm{t} \frac{3\cos^2 \theta - 1}{2} (3 m_{J'}^2 - 2)\right) + \mu B m_{J'},
\end{equation}
where $\Delta \alpha_\mathrm{s}$ is the scalar differential polarizability, $\alpha_\mathrm{t}$ is the tensor polarizability of the $\tP{1}$ state, $\mu = g_{J'} \mu_\mathrm{B}$ is the magnetic moment, and $\mu_\mathrm{B}$ is the Bohr magneton.
\begin{figure}[t!]
	\includegraphics{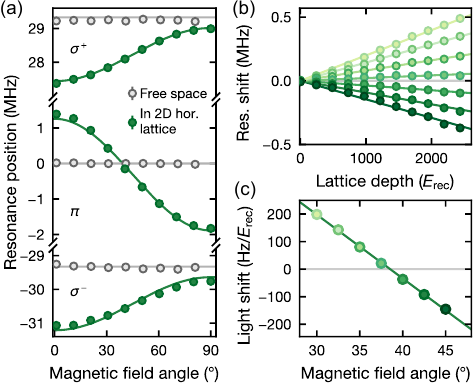}
	\caption{\label{fig:2}
		\textbf{Measuring the magic angle for the $\sS{0} \to \tP{1}$ transition at $759 \nm$}.
		$\left(\text{a}\right)$ Light shifted $\sS{0} \to \tP{1}$ $\sigma^+$, $\pi$, and $\sigma^-$ transitions (green) for various magnetic field angles relative to the vertical lattice polarization, and stationary free-space resonances (gray). The solid lines correspond to a single fit to the data using Eq.~\eqref{eq:magiccondition}. The error bars are obtained from the Lorentzian resonance fit uncertainty and are smaller than the datapoints.
		$\left(\text{b}\right)$ Resonance shifts of the $\pi$ transition for various magnetic field angles [color code as in (c)] and lattice depths close to the magic angle. The data is fitted with a linear function (solid lines).
		$\left(\text{c}\right)$ The fitted slopes are used to determine the magic angle using a linear fit since the curvature of the light shift in this regime is negligible.}
\end{figure}

\begin{figure*}[ht!]
	\includegraphics{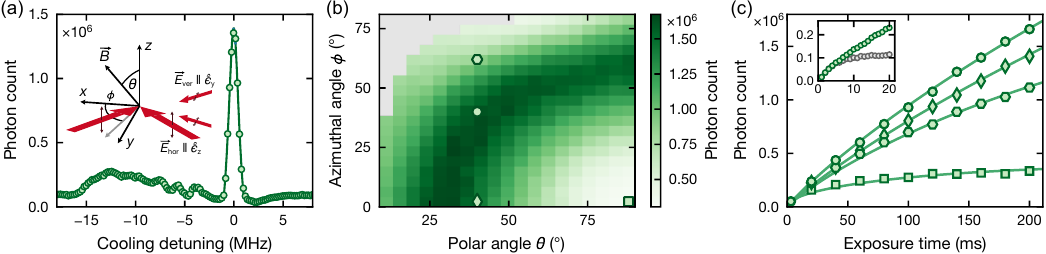}
	\caption{\label{fig:2.5}
		\textbf{Molasses cooling in a 3D lattice}.
		$\left(\text{a}\right)$ For a quasi-magic condition in a $\simeq \! 400 \uK$ deep 3D lattice we observe a $\simeq \! 600 \kHz$ wide molasses resonance close to the free-space $\tP{1}$ $\pi$ transition leading to a strongly enhanced collection of fluorescence photons after $200 \ms$ exposure time with weak $399 \nm$ probe light. In contrast, the light-shifted $\sigma^\pm$ transitions do not allow for efficient cooling of the whole cloud. The solid line is a Lorentzian fit to the cooling resonance data. Inset: Lattice beams (bright red single arrows), corresponding polarization vectors (dark red double-sided arrows), and magnetic field configuration.
		$\left(\text{b}\right)$ Cooling efficiency for different magnetic field orientations. While it is not possible to simultaneously reach the magic angle for the horizontal and vertical lattices, the resonance linewidth enables sufficiently quasi-magic conditions for a broad range of $B$-field angles.
		$\left(\text{c}\right)$ Evolution of the photon counts for varied exposure times. The only slightly curved trend observed for various magnetic field directions [indicated by the markers in $\left(\text{b}\right)$], demonstrates that the heating induced by scattering of $399 \nm$ light is at least partially counteracted. The solid lines serve as a guide to the eye. Inset: Comparison to an uncooled cloud (gray) that exhibits full atom loss after $\simeq \! 10 \ms$.}
\end{figure*}

To ascertain the magic angle in $\yb$, we perform atom loss spectroscopy on all three $m_{J'}$ transitions in a $\simeq \! 240 \uK \approx \! 2400 \, \Erec$ deep 2D lattice, where $\Erec = h^2 / (2m\lambda^2)$ is the lattice recoil energy with $h$ being Planck's constant, and $m$ the atomic mass.
This spectroscopy is performed for various magnetic field angles relative to the vertical polarization axis at a total field of $B \simeq \! 14 \Gauss$ and we compare the resulting resonance positions to the case of free-space resonances to determine the light shift according to Eq.~\eqref{eq:magiccondition} [Fig.~\subfigref{2}{a}].
The latter are determined within a short time-of-flight period during which the lattice is quenched off, while the direction and intensity of the circularly polarized spectroscopy beam is chosen such that the projection on the individual transitions is sufficiently large over the whole angle range and yields only minimally power-broadened linewidths, which we find to be the case for a beam approximately co-propagating with one of the horizontal lattice arms and $I \! \simeq \! I_\mathrm{sat}$.
This yields a magic angle for the $m_{J'} = 0$ state at $\simeq \! 35^\circ$ and further demonstrates the existence of a near-magic condition for the transitions to the $m_{J'} = \pm 1$ states for magnetic fields orthogonal to the lattice polarization.

To obtain a more precise result for the $m_{J'}\,=\,0$ magic angle, we scan the lattice depth for various magnetic field angles.
This results in linearly in- or decreasing resonance shifts as depicted in Fig.~\subfigref{2}{b}, such that we can use the fitted slopes to obtain the zero crossing of the total light shift at $\theta_\mathrm{magic} = 38.5(9)^\circ$ [Fig.~\subfigref{2}{c}].
Here, we make use of the vanishingly small curvature of the expected functional form in Eq.~\eqref{eq:magiccondition} and use a simple linear fit for this result.
While this fit yields a statistical uncertainty on the $\mathrm{mrad}$ level, the reported uncertainty is governed by systematic effects, where the dominant contribution stems from the calibration of $\theta$~\cite{SM}.

The magic angle facilitates fast efficient fluorescence imaging via the $\tP{1}$ state by directly collecting fluorescence photons scattered by the cooling light with a high-NA objective~\cite{lis:2023}.
In this work, we instead apply an additional weak probe beam of $399 \nm$ photons for imaging on the $\sP{1}$ state ($I \! \simeq \! 10^{-3} \, I_\mathrm{sat}$), which will be crucial for achieving single-site resolution in the retro-reflected clock-magic lattice with a small lattice constant of only $380 \nm$.
To counteract the resulting heating we perform molasses cooling with the horizontal MOT beams at an intensity of $I \! \simeq \! I_\mathrm{sat}$ and a detuning of $\delta \! \simeq \! -20 \kHz$ from the free-space resonance [Fig.~\ref{fig:2.5}].
In the 3D lattice geometry, we find an overall improved loading rate from the MOT for a horizontal polarization of the vertical lattice beams, which we attribute to a larger interference contrast and, thus, a stronger vertical confinement. 
As a result, a simultaneous magic condition for all three lattice beams cannot be achieved. 
Nonetheless, we observe that we can reach a quasi-magic condition for all three lattice beams by also adjusting the azimuthal magnetic field angle $\phi$ [Fig.~\subfigref{2.5}{b}].
This way, we find a contour on which the total number of photon counts for a constant exposure time remains close to the maximum, indicating efficient cooling conditions.
\begin{figure}[t!]
	\includegraphics{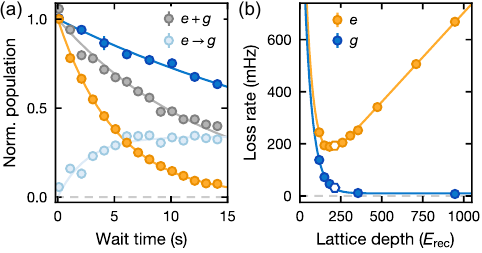}
	\caption{\label{fig:3}
		\textbf{$\tP{0}$ clock-state lifetime}.
        $\left(\text{a}\right)$ Lifetime of sideband-cooled atoms in a $220 \, \Erec$ deep 2D lattice initialized in the $\sS{0}$ state (dark blue) or $\tP{0}$ state (orange).
        Raman-scattering-induced losses out of $\tP{0}$ partially reappear in the $\sS{0}$ state and can be detected from the total atom number (gray) via the difference to the $\tP{0}$ decay curve, yielding the number of converted atoms (light blue).
        The data is fitted with solutions to coupled differential equations (solid lines), extracting the lifetimes of each state and the fraction of atoms converted through the aforementioned Raman channel~\cite{siegel:2024}.
        Error bars correspond to the standard error of the mean of three averages.
		$\left(\text{b}\right)$ While the ground-state loss rate benefits from deep lattices, Raman losses lead to a linear increase in the $\tP{0}$ clock-state loss rate for lattice depths $\gtrsim \! 250 \, \Erec$.
        Below the optimal trap depth of $220 \, \Erec$ (white hexagon), strong inelastic collisions and spilling losses start to become dominant.
        Error bars from the exponential fit uncertainties are smaller than the datapoints, and the solid lines are exponential fits ($\sS{0}$) in combination with a linear term ($\tP{0}$).}
\end{figure}
The contour can be understood as the light shift introduced by the horizontal lattice being approximately canceled by an opposite light shift from the vertical lattice, with additional effects stemming from the sizeable electric field splitting compared to the modest external magnetic field strength of $1.3 \Gauss$~\cite{SM}.
\begin{figure*}[ht!]
	\includegraphics{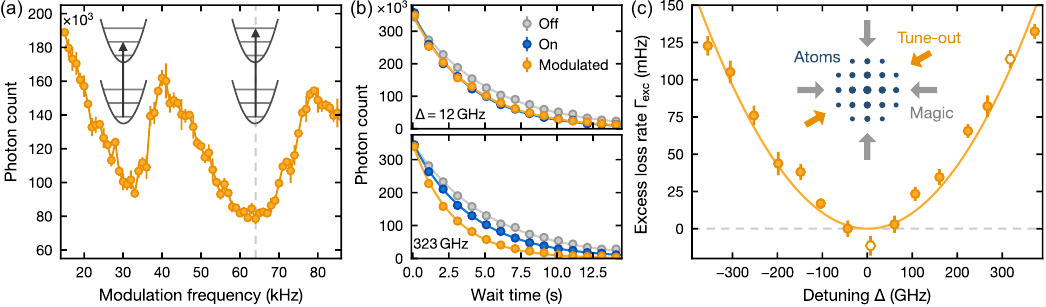}
	\caption{\label{fig:4}
		\textbf{Determination of the $\tP{0}$ clock-state tune-out wavelength}.
		$\left(\text{a}\right)$ Modulation spectroscopy of clock-state atoms $\simeq \! 1 \THz$ detuned from the tune-out wavelength. After $1.5 \s$ of amplitude modulation strong loss features can be found at the longitudinal trap frequency of $\simeq \! 32 \kHz$ and its first multiple, with the former corresponding to excitations by one and the latter by two harmonic oscillator quanta (insets). The solid line is a linear interpolation.
		$\left(\text{b}\right)$ Clock-state lifetimes in the bare clock-magic lattice (gray), with the additional tune-out lattice at constant amplitude (blue), and with amplitude modulation at $f_\mathrm{mod} = 64 \kHz$ [orange, gray dotted line in $\left(\text{a}\right)$]. While the presence of the tune-out lattice induces a moderate, constant excess loss, the modulation leads to pronounced losses for large detunings from the tune-out wavelength (bottom panel) and does not affect the lifetime close to it (top panel). We fit the data with single exponential functions (solid lines). The error bars obtained from three averages are smaller than the datapoints.
		$\left(\text{c}\right)$ Excess loss rate from modulation compared to the static tune-out lattice case for varying laser frequencies [white markers indicate the parameters used for the measurements in (b)]. The data follows a quadratic trend (solid line). The error bars are obtained from the standard deviation of the individual lifetime fit uncertainties. Inset: Sketch of the total lattice geometry, with the tune-out lattice at an arbitrarily chosen angle of $\simeq \! 30^\circ$ with respect to the magic 2D lattice.}
\end{figure*}
We further note that even for a subpar field orientation a large fraction of atoms remains cooled, whereas the photon count without cooling already saturates at $\simeq \! 10 \ms$ due to complete scattering-induced atom loss, as can be seen in Fig.~\subfigref{2.5}{c}, inset.
Using an exposure time of $200 \ms$, we can thus collect $16\times$ more fluorescence photons per atom on the camera than without cooling, allowing us to resolve atom number variations on the order of $100$ atoms.

To study the lifetime of atoms in the $\tP{0}$ state, we apply a resonant, $\simeq \! 260 \us$ long clock $\pi$-pulse, followed by a strong $399 \nm$ pulse to remove all atoms that were not transferred to the metastable state and a repump pulse before starting the fluorescence imaging sequence~\cite{SM}.
By varying the wait time between the clock-excitation and repump pulse we can therefore trace the $\tP{0}$ loss rate.
In contrast to the ground state, where vacuum losses are typically predominant, off-resonant Raman scattering of $759 \nm$ trap photons limits the $\tP{0}$ lifetime in deep traps~\cite{siegel:2024}.
Hence, the tune-out measurement has to be performed in a weak lattice to achieve the best resolution for the heating-induced excess loss rate measurements.
To mitigate spilling and two-body inelastic collisions due to atoms occupying higher vibrational bands that can tunnel fast in a shallow optical lattice, we perform cooling prior to the measurement scheme described above. 
Since this work employs spinless, bosonic \yb, cooling techniques that leverage the existence of multiple ground states, such as $\Lambda$-enhanced gray molasses or Raman sideband cooling~\cite{angonga:2022,grier:2013,kaufman:2012,thompson:2013,hamann:1998,kermanphd:2002}, are out of reach.
Instead, we utilize resolved sideband cooling on the clock transition similar to Ref.~\cite{nemitz:2016}, which we adapt for use in a 2D lattice~\cite{SM,kroeze:2024}.
We note that horizontal cooling in a pure 2D lattice is sufficient to localize and separate the atoms well, while the third direction can remain uncooled. 
After clock-sideband cooling we obtain a mean motional occupation number of $\bar{n}\simeq \! 0.1$ along each horizontal direction in a $\simeq \! 220 \, \Erec$ deep lattice.
This enhances the $\tP{0}$-state lifetime to $\tau = 5.2(1) \s$, which is defined as the 1/e-decay time extracted with an exponential fit [Fig.~\subfigref{3}{a}].
This lifetime is still limited by Raman scattering as evidenced by the substantially shorter lifetime of the $\tP{0}$ state compared to that of $\sS{0}$, and the crossover to a linear increase of the $\tP{0}$ loss rate for deep lattices [Fig.~\subfigref{3}{b}].
For shallow lattice depths, strong inelastic collisions and spilling losses increase the loss rate of both $\sS{0}$ and $\tP{0}$ states.
A fraction of Raman-scattered atoms from the $\tP{0}$ state reappears in the $\sS{0}$ ground state by decaying via the intermediate $\tP{1}$ state, while atoms scattered to the $\tP{2}$ state are anti-trapped and subsequently lost.
The former pathway can be detected by measuring the total atom number, i.e., by omitting from the detection sequence a dedicated resonant pulse that normally removes any $\sS{0}$ atoms.
The observed fraction is consistent with the results reported in Ref.~\cite{siegel:2024}.

As a next step, while the magic lattice remains in a 2D geometry, superimposing a retro-reflected lattice close to the tune-out wavelength at an arbitrarily chosen angle of $\simeq \! 30^\circ$ to the 2D lattice enables us to probe sinusoidal amplitude-modulation-induced parametric heating.
Similar to the 1D lattice case utilized in previous experiments~\cite{hoehn:2023,heinz:2020}, the incommensurability of wavelengths then incurs mostly phase modulation for atoms in certain lattice sites, while others predominantly experience amplitude modulation as shown in Fig.~\subfigref{4}{a}, where the modulation spectrum displays two strong resonances at the longitudinal trap frequency and its first multiple.
We note that a pure phase modulation from a running wave induced by two detuned beams would provide a stronger single resonance, however, at the expense of a four times lower peak intensity.
Choosing the modulation frequency of $64 \kHz$ that shows the strongest loss response, we study the effect of the tune-out lattice on the clock-state lifetime as a function of the laser frequency [Fig.~\subfigref{4}{b}].
As can be expected from the negligibly small curvature of the $\tP{0}$ polarizability within the tuning range and from simple perturbation theory calculations~\cite{heinz:2020,savard:1997}, the modulation-induced excess loss rate scales quadratically around the tune-out frequency [Fig.~\subfigref{4}{c}].
However, this simple model breaks down for very fast loss dynamics.
Here, we observe the emergence of additional, slower timescales arising in the lifetime curves, where the separation into mostly phase- and amplitude-modulated lattice sites and the strong dependence of tunneling rates on the harmonic oscillator state starts to play a role.
We therefore restrict the analysis to data that is well described by a single exponential function and perform each measurement at three different average tune-out lattice powers~\cite{SM}.
Computing the weighted mean of the resulting minima then yields a $\tP{0}$ tune-out frequency of
\begin{equation}
	f_\mathrm{to} = 519.9199 \pm 29_\mathrm{stat} \left(^{+57}_{-45}\right)_\mathrm{sys} \THz.
\end{equation}
Here, the statistical uncertainty corresponds to the standard error of the mean, while for the systematic uncertainty we take the deviation from the perturbative regime into account by fitting both the limited and the full dataset with an empirical function that includes a smooth crossover to a linear regime for large detunings~\cite{SM}.

We can compare the measured tune-out wavelength value to the empirical model presented in Ref.~\cite{hoehn:2023} and find excellent agreement.
Including $f_\mathrm{to}$ in the model therefore leads to marginal changes of the fit parameters~\cite{SM}.
Leveraging the vanishing $\tP{0}$ polarizability at $f_\mathrm{to}$, we further measure the ac Stark shift of the ground state by means of high-resolution clock spectroscopy in the presence of a static, unreflected dipole beam at the tune-out wavelength for varying powers.
Repeating this measurement after tuning the laser to the clock transition enables a direct comparison of the relative light shift ratio, which we quantify to be $\Delta\alpha_\mathrm{to} / \Delta\alpha_\mathrm{clock} = 1.014(11)$.
Together with the previously determined clock probe shift of $15(3) \Hz / (\mathrm{W}\!/\mathrm{cm}^2)$ reported in Ref.~\cite{poli:2008}, we thus find the empirical model's prediction for the $\sS{0}$ polarizability at the tune-out wavelength of $V\!_\mathrm{ac}/I=-13.0 \, h \Hz / (\mathrm{W}\!/\mathrm{cm}^2)$ to be well within the error bar.

Notably, the presence of the tune-out beam leads to enhanced losses from off-resonant Raman processes, which however do not vary over the chosen detuning range and thus yield a constant offset to the lifetime in the bare 2D lattice.
Utilizing the light shift measurement reported above as an intensity calibration~\cite{SM}, we measure the Raman loss rate to be $12(3) \times 10^{-6} \Hz / (\mathrm{W}\!/\mathrm{cm}^2)$, which reasonably agrees with a theoretical estimate of $10.2 \times 10^{-6} \Hz / (\mathrm{W}\!/\mathrm{cm}^2)$~\cite{siegel:2024,SM}.

\section{Summary and conclusion}
In this work we have presented the first measurement of a metastable excited-state tune-out wavelength, aided by a combination of sideband cooling on the $\sS{0} \to \tP{0}$ and molasses cooling on the $\sS{0} \to \tP{1}$ transition.
For the latter, we have identified a novel magic angle that allows for triple-magic operation at $759 \nm$ trap light with bosonic $\yb$ atoms and characterized a stability region of effective cooling during fluorescence imaging in a 3D clock-magic lattice.
This represents a major step towards the realization of an Yb quantum gas microscope with a clock-magic lattice, while the newly determined tune-out wavelength greatly enriches the toolbox for applications in quantum computing and simulation.
In particular, tune-out wavelengths may be beneficial for rearrangement of atoms in dense arrays or lattices.
Here, one can utilize the ability to apply local light shifts to selectively shelve atoms to the metastable clock state with minimal impact on their coherence as well as to shuttle ground-state atoms to the desired final position without disturbing the potential of shelved atoms, enhancing the scalability of existing quantum computation schemes~\cite{bluvstein:2024,lis:2023,norcia:2024}. Moreover, the triple-magic condition for $\tP{1}$ and $\tP{0}$ in a short-spacing retro-reflected lattice offers exciting opportunities for the realization of novel light-matter interfaces in extreme sub-wavelength arrays~\cite{henriet:2019,bekenstein:2020,masson:2024}.

\begin{acknowledgments}
We thank Etienne Staub  for experimental assistance. This project has received funding from the Deutsche Forschungsgemeinschaft (DFG, German Research Foundation) under Germany’s Excellence Strategy -- EXC-2111 -- 390814868 and via Research Unit FOR5522 under project number 499180199. We further acknowledge funding from the European Research Council (ERC) under the European Union’s Horizon 2020 research and innovation program (grant agreement No. 803047), from the German Federal Ministry of Education and Research (BMBF) via the funding program quantum technologies -- from basic research to market (contract number 13N15895 FermiQP) and from the Initiative \textit{Munich Quantum Valley} from the State Ministry for Science and the Arts as part of the High-Tech Agenda Plus of the Bavarian State Government. This work has further received funding under Horizon Europe programme HORIZON-CL4-2022-QUANTUM-02-SGA via the project 101113690 (PASQuanS2.1) and from the European Union’s Horizon 2020 research and innovation programme under Grant Agreement No 101017733 (DYNAMITE, DFG project number 499183856). 
\end{acknowledgments}


%

\cleardoublepage
\section*{Supplemental Material}

\renewcommand{\thefigure}{S\arabic{figure}}
\renewcommand{\theHfigure}{S\arabic{figure}}
\renewcommand{\theequation}{S.\arabic{equation}}
\renewcommand{\thesection}{S.\Roman{section}}
\renewcommand{\thetable}{S\arabic{table}}
\setcounter{figure}{0}
\setcounter{equation}{0}
\setcounter{section}{0}
\setcounter{table}{0}

\section{Experimental sequence}\label{sec:expseq}
\paragraph{Lattice loading:} 
The experiment starts by loading a 3D MOT on the $\sS{0} \rightarrow \tP{1}$ transition for $500 \ms$, which typically results in a total atom number of $\simeq \! 4 \times 10^6$.
We note that compared to the MOT setup described in Ref.~\cite{hoehn:2023}, we inserted a high-resolution objective (NA $= 0.7$), where the bottom MOT beam is now projected onto the back focal plane of the objective.
At the beginning of the MOT compression stage, we quench on the 3D clock-magic lattice to $\simeq \! 330 \, \Erec$ in each horizontal direction and to $\simeq \! 720 \, \Erec$ in the vertical axis. 
After $ 50 \ms$ of equilibration time in the lattice while keeping the MOT on, both the MOT beams and the magnetic field gradient are extinguished, and after an additional $25 \ms$ of wait time the vertical lattice is ramped down within $50 \ms$. 
\paragraph{Magic-angle measurement in 2D lattice:} 
For the magic-angle measurements in Fig.~2 of the main text, the 2D lattice is quickly ramped to the desired depth.
Spectroscopy is performed using a $500 \us$ long pulse on the $\tP{1}$ transition, where the light is applied via one of the horizontal MOT beams.
\paragraph{Free-space spectroscopy:}
For the free-space spectroscopy in Fig.~2 of the main text, the 2D lattice is fully quenched off before applying the $500 \us$ long spectroscopy pulse and then quenched back on, retaining more than $70\%$ of the atoms.
\paragraph{Lifetime measurements:}
For the $\tP{0}$ state lifetime presented in Fig.~4 in the main text, we first perform clock-sideband cooling in the 2D lattice, as described in Section~\ref{sec:cooling}.
Subsequently, the horizontal lattices are ramped to $110 \, \Erec$ depth along each direction, followed by resonant clock excitation.
Any remnant ground-state atoms are subsequently removed by means of a $3 \ms$, resonant pulse on the $\sP{1}$ transition, followed by a variable wait time before imaging.
The $\sS{0}$ lifetime is measured identically but by omitting the excitation and removal pulses.
\paragraph{Tune-out measurement:}
For the tune-out measurements in Fig.~5 in the main text, the sequence follows that of the clock state lifetime, with the following modifications.
After clock excitation and ground-state atom removal, the tune-out lattice is ramped to the desired intensity over $5 \ms$.
Then, the intensity of the tune-out lattice is either held constant or modulated sinusoidally, for a variable wait time.
After the wait, the tune-out lattice is ramped down over $5 \ms$ prior to imaging.
\paragraph{Imaging:}
When imaging $\tP{0}$ atoms, we first remove any $\sS{0}$ atoms that may have appeared as a result of Raman scattering, identical to how these are removed after clock excitation.
Then, all three lattices are ramped up to their maximal depth within $5 \ms$.
We then apply a repump pulse to transfer the clock-state atoms back to the ground state, where they are imaged on the $\sP{1}$ transition while being molasses-cooled using the MOT beams as described in the main text.
Imaging of $\sS{0}$ atoms proceeds identically but by omitting the removal and repump pulses.

\section{Empirical polarizability model}\label{sec:polarizability}
Utilizing the empirical model for the $\sS{0}$ and $\tP{0}$ ac polarizabilities developed in our previous work \cite{hoehn:2023}, we obtain only marginal corrections from $\lambda_\mathrm{eff,\tP{0}} = 376.1 \nm$ to $374.7 \nm$ and from $\Gamma\!_\mathrm{eff,\tP{0}} = 22.9 \MHz$ to $23.6 \MHz$ upon inclusion of the newly measured tune-out wavelength at $576.6 \nm$.
This does not lead to visible changes in the polarizability curves, and the benchmark polarizability ratios at $670 \nm$, $671.5 \nm$, and $690.1 \nm$ are affected on a level well below the measurement uncertainty~\cite{riegger:2018,darkwahoppong:2022}.

Due to the increasing relevance of state-dependent potentials also for the intercombination line ($\sS{0}\rightarrow\tP{1}$) and a multitude of measured magic wavelengths at specific magic angles \cite{yamamoto:2016,jenkins:2022,ma:2022,lis:2023,norcia:2023,norcia:2024}, we extend the polarizability model to the $\tP{1}$ state.
We use the general expression for the total light shift~\cite{steck}
\begin{align}\label{eq:totalshift}
	V\!_\mathrm{ac}(\omega) = -\frac{I}{2c\epsilon_0} & \left[\alpha^{(0)}(\omega) \right. \\
	+ & \alpha^{(1)}(\omega) q(\hat{k} \cdot \hat{z}) \frac{m_F}{F} \nonumber\\
	+ & \left. \alpha^{(2)}(\omega) \frac{3 \lvert \hat{\varepsilon}_z \rvert^2 -1}{2} \frac{3m_F^2 - F(F+1)}{F(2F-1)} \right], \nonumber
\end{align}
with the scalar, vector, and tensor polarizabilities denoted as $\alpha^{(0,1,2)}$, respectively, the total atomic angular momentum $F$, the angular frequency of the trapping light $\omega$, its intensity $I$, unit wave vector $\hat{k}$, polarization vector $\hat{\varepsilon}$, quantization axis $\hat{z}$, and the degree of circular polarization $q$.
Due to the finite electronic angular momentum, scalar and tensor light shifts are typically comparably strong, while the vector shift vanishes for linearly polarized trapping light ($q = 0$) as is the case in our experiment.
The scalar and tensor shifts can be further expressed in terms of the reduced dipole matrix element as~\cite{steck}
\begin{equation}
	\alpha^{(0)}(\omega) = \sum_{F'} \frac{2 \omega_{F'F}}{3 \hbar (\omega_{F'F}^2 - \omega^2)} \lvert \langle F |\!| \mathbf{d} |\!| F' \rangle \rvert^2
\end{equation}
and
\begin{align}
	\alpha^{(2)}(\omega) = \sum_{F'} (-1)^{F'+F} \sqrt{\frac{40F(2F+1)(2F-1)}{3(F+1)(2F+3)}} \times \\
	 \begin{Bmatrix}
		1 & 1 & 2 \\
		F & F & F'
	\end{Bmatrix} \frac{\omega_{F'F}}{\hbar(\omega_{F'F}^2 - \omega^2)} \lvert \langle F |\!| \mathbf{d} |\!| F' \rangle \rvert^2, \nonumber
\end{align}
which can in turn be linked to the transition linewidths \cite{hoehn:2023}.

\begin{figure}[t!]
	\includegraphics{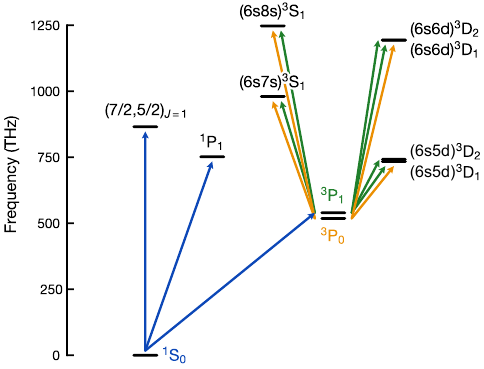}
	\caption{\label{fig:levels}
		\textbf{Energy levels and transitions used for the empirical polarizability models}.
		Following the treatment of the $\sS{0}$ and $\tP{0}$ ac~polarizabilities in Ref.~\cite{hoehn:2023}, we use a limited dataset of relevant states to describe the polarizability of the $\tP{1}$ state in the optical spectrum. In addition to the $\tS{1}$ and $\tD{1}$ states that are strongly coupled to $\tP{0}$, in the case of $\tP{1}$ also transitions to $\tD{2}$ states need to be taken into account.}
\end{figure}
As with the model for the $\tP{0}$ polarizability, we truncate the set of transitions that are included directly, as shown in Fig.~\ref{fig:levels}, and absorb the neglected higher-lying transitions in an effective line at $\lambda_\mathrm{eff,\tP{1}}$ with width $\Gamma\!_\mathrm{eff,\tP{1}}$. 
By accounting for the fine-structure splitting, we directly relate $\lambda_\mathrm{eff,\tP{1}}$ to $\lambda_\mathrm{eff,\tP{0}}$, yielding $\lambda_\mathrm{eff,\tP{1}}=384.9\nm$ and leaving only $\Gamma\!_\mathrm{eff}$ as a free parameter.
Furthermore, we take the finite transition strengths to $\tD{2}$ states into account by splitting the linewidth of the effective excited state into two components with $J=1$ and $J=2$, respectively.
Their corresponding strengths are deduced from the nearby $(6s7d)\tD{1}$ and $\tD{2}$ transition strength ratio as a simplification in order to minimize the number of free parameters.
Similarly, the identical resonance frequency can be motivated by the very small fine-structure splitting of the $(6s7d)\tD{}$ states.
Using the magic wavelengths that have been determined for $\ybo$ \cite{ma:2022,lis:2023,norcia:2023,norcia:2024} and our magic angle at $759 \nm$ for $\yb$, we obtain high agreement with our empirical model for a combined effective linewidth of $\Gamma\!_\mathrm{eff, \tP{1}} \simeq \! 2\pi \times 59 \MHz$. We note that the measured scalar and tensor polarizabilities at $532 \nm$ in Ref.~\cite{yamamoto:2016} cannot be reproduced by our model.

\section{Sideband cooling}\label{sec:cooling}
As described in the main text, we cool the atomic sample to prolong the excited state lifetime. Lower temperatures allow us to use less lattice light to confine the atoms, thereby minimizing the Raman scattering losses discussed in the main text and Sec.~\ref{sec:Raman}. At the same time, reduced temperatures suppress inelastic losses between pairs of excited state atoms.
In this Section we briefly outline the general performance and sequence, details on the adaptation of this standard technique to the 2D lattice will be described elsewhere~\cite{kroeze:2024}.

To perform cooling, atoms are excited on the resolved red sideband of the ultranarrow clock transition $\sS{0}\to\tP{0}$~\cite{nemitz:2016}. 
The cooling cycle is closed by repumping on the $\tP{0}\to\tD{1}$ transition, which decays via $\tP{1}$ back to the ground state $\sS{0}$, as described in the main text and \ref{sec:expseq}. We choose this repumping transition because of the reasonably short radiative lifetime of $\simeq\! 300\ns $, the favorable branching ratios with only marginal, $\simeq 3 \%$ losses into the dark and anti-trapped \tP{2} state, and the long wavelength of $1389\nm$, resulting in a low recoil energy of $h \times 595\Hz$.
\begin{figure}[t!]
	\includegraphics{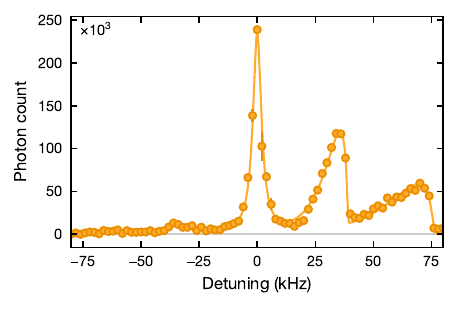}
	\caption{\label{fig:cooling}
		\textbf{Sideband spectroscopy after clock-sideband cooling in a 2D lattice}.
        Resolved sideband spectrum in a balanced, 2D lattice with $\simeq220 \Erec$ combined depth. The figure shows the first two red and blue sidebands, where the second red sideband is fully suppressed and a small fraction of the first red sideband is still visible. Error bars correspond to the standard error of the mean over three averages. The solid line is a guide to the eye. 
		}
\end{figure} 

Cooling is performed in a 2D lattice configuration, where the two lattices are matched in depth to $\simeq\! 330\, \Erec$ each. 
Two clock beams, each (approximately) co-propagating with one of the lattices, are used to drive separate pulses on the red sideband transition. 
The sideband pulse duration is limited by the achievable clock Rabi rate for which we use a strong magnetic field of $\simeq\! 400 \Gauss$ to induce the clock transition~\cite{taichenachev:2006}. 
Together with $\sim\!100\mW$ of power, each beam reaches a carrier transition Rabi rate of $\Omega\simeq2\pi\times 3.2\kHz$. 
We send 15 clock cooling pulses with $1 \ms$ duration, each followed by a $500 \us$ repumping pulse. To efficiently address all atoms in the 2D lattice, the frequency of the clock beams is swept during this time from $-30 \kHz$ to $-72 \kHz$ to address inhomogeneities in the sideband frequency~\cite{kroeze:2024}. 
This sequence is then repeated along the other direction to complete one cooling ``cycle'', and we perform 10 such cooling cycles.
Prior to cooling, time-of-flight expansion reveals the atoms to have a temperature of $T\!_\mathrm{hor} \simeq10\uK$ ($\bar{n}\simeq 2.6$).
After cooling, this temperature is reduced to $T\!_\mathrm{hor} \simeq \! 1.6\uK$ ($\bar{n}\simeq 0.15$) as determined using sideband spectroscopy. 
The spectroscopy utilizes a 30 ms long clock pulse with $\Omega\simeq2\pi\times 2.1\kHz$, ensuring that the transition is incoherently driven.
Figure~\ref{fig:cooling} shows the sideband spectrum after the subsequent adiabatic ramp of the lattice depth to that yielding maximal lifetime, i.e.\ $\simeq\!220\,\Erec$.
In that lattice, the thermal occupation results in an average hopping rate of $\simeq\! 150 \mHz$ compared to $\simeq\! 170\Hz$ prior to cooling.

We note that the vertical direction is not explicitly cooled.
After application of a coherent clock $\pi$-pulse we observe rapid ($\simeq \! 5 \ms$) losses of $\sim40\%$ of the $\tP{0}$ atoms. 
We ascribe this to inelastic collisions between multiple $\tP{0}$ atoms residing on a single lattice site facilitated by the high temperature along the weakly confined vertical direction. 
Indeed, we confirm that a larger particle density results in more severe losses. 
Adding vertical confinement inhibits these losses, though at the cost of increased Raman scattering, and we opt to accept this rapid initial loss in favor of the longer lifetime.
To avoid any effects from these losses affecting the lifetime measurement, we thus choose a minimal wait time of $100 \ms$ after the application of the clock pulse before we start to record the lifetime.
This further ensures that any detrimental effects of the settling of the strong external magnetic field, observed to take $\simeq \! 20 \ms$,  are avoided.

\section{Intensity calibration}\label{sec:intensity-calibration}
Determining the intensity of the tune-out beam is a crucial ingredient for an accurate measurement of the Raman scattering rate (Section~\ref{sec:Raman}). To this end, we tune the laser to be resonant with the clock transition and illuminate the atomic sample with a dipole beam (retro-reflection is blocked). 
Moreover, we work at low optical power $P \leq 10 \mW$ to limit power broadening.
This way, we measure an ac Stark shift response of $98.0(6) \kHz / \mathrm{W}$. Utilizing the experimentally determined polarizability reported in Ref.~\cite{poli:2008} and the carefully calibrated total power of the dipole beam, we obtain a waist of $99(10) \um$.
We then set the laser frequency to the tune-out value and repeat the spectroscopy measurements using the same settings. We obtain a response of $99.3(8) \kHz / \mathrm{W}$, which we use to compute the relative light shift of $\Delta\alpha_\mathrm{to} / \Delta\alpha_\mathrm{clock} = 1.014(11)$ as mentioned in the main text.

\section{Raman scattering}\label{sec:Raman}
Both the magic wavelength lattice and the tune-out light induce Raman scattering which results in additional loss of \tP{0} atoms.
\begin{figure*}[t!]
	\includegraphics{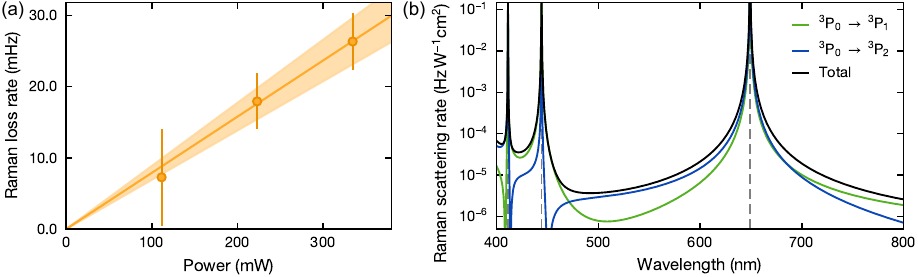}
	\caption{\label{fig:Raman}
		\textbf{Raman loss rate at the tune-out wavelength and in the visible spectrum}.
		$\left(\text{a}\right)$ The clock-state lifetime is reduced by the presence of the tune-out beam with a linear dependence of the Raman loss rate on the power. 
        The error bars are obtained from the standard deviation of the individual lifetime fit uncertainties.
        The solid line is a weighted linear fit to the data, and the shaded area depicts the fit uncertainty.
		$\left(\text{b}\right)$ Theoretical estimate of the intensity-dependent Raman scattering rate by generalizing the contributions of relevant transitions in Ref.~\cite{siegel:2024}.}
\end{figure*}
For the wavelengths and intensities employed here, the Raman scattering mechanism dominates over other loss mechanisms such as vacuum-limited collisions, spontaneous emission and black body radiation. 
Raman scattering out of the \tP{0} state predominantly results in decay to the other fine-structure states, i.e., \tP{1} and \tP{2}.
While atoms in the \tP{1} state quickly decay back to the ground state \sS{0} and can be detected as shown in Fig.~4 in the main text, the \tP{2} state is long-lived and anti-trapped by the clock-magic lattice, resulting in loss of the \tP{2} population.

In this Section we provide a quantitative analysis of the observed Raman scattering and details of the theoretical calculation of the expected scattering rates.

\subsection{Experiment}
As shown in Fig.~5 of the main text, the presence of a constant, i.e.\ non-modulated, tune-out lattice results in excess loss of atoms out of the \tP{0} state. Fig.~\subfigref{tuneout-systematics}{d} shows that this excess loss is independent of laser frequency over the range explored in this work, and that it is linearly proportional to the intensity as expected for Raman scattering. To quantitatively analyze the Raman scattering induced by the tune-out light, we block the retro-reflecting mirror resulting in a dipole trap rather than an optical lattice. This simplifies the analysis as systematic effects such as imperfect lattice alignment are eliminated.

The excess loss rate observed in this configuration is shown in Fig.~\subfigref{Raman}{a}. 
Together with the intensity calibration described in Sec.~\ref{sec:intensity-calibration}, we can accurately extract the relation between Raman loss rate and light shift, yielding ${8(1)\times10^{-7} \Hz_\mathrm{loss} / \Hz_\mathrm{light shift}}$. 
While this quantity is practically useful, it does not enable comparison to theoretical calculations. 
For that, we use the beam waist inferred from the light shift spectroscopy and find $12(3)\times10^{-6} \Hz / (\mathrm{W}\!/\mathrm{cm}^2)$. 
We note that again the increased uncertainty is dominated by the measurement of the probe light shift of resonant clock light reported in Ref.~\cite{poli:2008}.

\subsection{Theory}
This type of off-resonant scattering rate can be calculated using the Kramers-Heisenberg formula~\cite{bransden:2003}. Assuming a linearly polarized lattice and neglecting the hyperfine interaction, the total scattering rate $\Gamma_{i\rightarrow f}(\omega)$ from initial state $i$ to final state $f$ induced by radiation of angular frequency $\omega$ and intensity $I$ can be written as~\cite{siegel:2024}
\begin{equation}
	\Gamma_{i\rightarrow f}(\omega) = \Lambda_{if}\frac{(\omega - \omega_{fi})^3}{6\pi\epsilon_0^2c^4\hbar}|\alpha_{if}(\omega)|^2I,
\end{equation}
where the branching ratio (after summing over the hyperfine substates) $\Lambda_{if}=1$ for all our transitions of interest, $\hbar\omega_{fi}$ is the splitting between states $i$ and $f$, and
\begin{equation}
	\alpha_{if}(\omega) = \sum\limits_k \alpha_{ikf}(\omega)
\end{equation}
sums the contributions over all intermediate states $k$. We note that while Ref.~\cite{siegel:2024} expresses the scattering rate in terms of trap depth, we instead write it explicitly in terms of the intensity. This is to avoid confusion at tune-out wavelengths, where the former vanishes but the latter remains well-defined.

These calculations, in particular of $\alpha_{ikf}(\omega)$, require accurate knowledge of not only transition frequencies to higher excited states, but also dipole matrix elements. 
While our empirical polarizability model described in Sec.~\ref{sec:polarizability} sufficiently captures the relevant transitions out of \sS{0}, \tP{0}, and \tP{1}, we have not developed a similar model for \tP{2}. 
Instead, we utilize the results of the calculations in Ref.~\cite{siegel:2024} and generalize these to other wavelengths of interest. 
Specifically, Table~I of the Supplemental Material of Ref.~\cite{siegel:2024} reports the quantities $\alpha_{ikf}(\omega_\text{magic})$ for $\omega_\text{magic}=2\pi c/\lambda$ with $\lambda=759.3\,$nm. These can be converted to other frequencies using
\begin{equation}\begin{split}
	\frac{\alpha_{ikf}(\omega')}{\alpha_{ikf}(\omega)} = &\left[\frac{1}{\omega_{ki}-\omega'}+\frac{(-1)^{J_f}}{\omega_{ki}+\omega_{if}+\omega'}\right]\\
&\times\left[\frac{1}{\omega_{ki}-\omega}+\frac{(-1)^{J_f}}{\omega_{ki}+\omega_{if}+\omega}\right]^{-1},
\end{split}\end{equation}
where we used $\omega_{kf} = \omega_{ki} + \omega_{if}$. 
This conversion depends solely on the two frequencies $\omega$ and $\omega'$, the final state angular momentum $J_f=1(2)$ for \tP{1}(\tP{2}), the well-known fine-structure splittings $\omega_{if}$ and the excitation frequencies $\omega_{ki}$. 
The latter is unknown (or undefined) for the contributions from ``all other valence'' as well as the core-excited contribution. 
But since these are high energy excitations, we approximate $\omega_{ki}\gg\omega,\omega'$ which means that these contributions are independent of wavelength. 
This assumption breaks down for short wavelengths, but since their relative contributions are small we nonetheless assume validity for wavelengths $\gtrsim 350\nm$. 
Finally, we note that while the results in Ref.~\cite{siegel:2024} are listed for \ybo, we perform our calculations for \yb. 
In doing so, we ignore any isotope effects to the reduced dipole matrix elements, i.e.\ intermediate state lifetimes. 
Isotope shifts to the fine-structure splittings $\omega_{if}$ and the excitation frequencies $\omega_{ki}$ are accounted for. 
Regardless, we estimate that isotope effects have negligible contribution. 
As a consistency check, we use the same approach to calculate the polarizability of \tP{0}, $\alpha(\omega) = \alpha_{ii}(\omega)$, for general frequencies $\omega$. 
A comparison to our independent calculation described in Sec.~\ref{sec:polarizability} shows good agreement except very close to resonant transitions out of \tP{0}, where details such as isotope shifts become more important. 
We believe this does not impact the calculations of the Raman scattering rates of interest.

Fig.~\subfigref{Raman}{b} shows the Raman scattering rate for the two relevant two-photon transitions as a function of wavelength. 
As expected, divergences appear at resonances out of the \tP{0} state, whose locations are indicated by the dashed lines. 
The arrows indicate the two wavelengths of interest. 
At the $759\nm$ magic wavelength, our calculation reproduces the results of Ref.~\cite{siegel:2024} (by construction, since no frequency conversion is needed here). 
There, the total scattering rate is $\Gamma_{\tP{0}\rightarrow\tP{1}}+\Gamma_{\tP{0}\rightarrow\tP{2}}=4.40 \times 10^{-6} \Hz / (\mathrm{W}\!/\mathrm{cm}^2)$, with 65\% of the scattering occurring to the \tP{1} state. 
At the tune-out wavelength, the total scattering rate is $10.2 \times 10^{-6} \Hz / (\mathrm{W}\!/\mathrm{cm}^2)$, which is the figure reported in the main text. 
We note that here the dominant scattering channel is reversed, with only 26\% scattering to \tP{1}.

\section{Systematic uncertainties in the magic angle measurements}
Various error sources factor into the systematic uncertainty of the magic angle measurements, which we now discuss.
Firstly we consider the uncertainty in the calibration between the current applied to bias coils and the resulting magnetic field. 
The bias coils consist of three separate coil pairs along $x$, $y$ and $z$ respectively.
Each coil pair is calibrated via free-space spectroscopy of the $\tP{1}$ Zeeman states, and results in relative uncertainties of $0.26\%$ for the vertically oriented coils and $0.04\%$ for the horizontal coils along $y$.
Notably, by tracing the $\tP{1}$ $m_{J'}=1$ resonance for various magnetic field angles we also find a small deviation from a perfectly orthogonal relative orientation between the fields created by these coils.
The extracted angle of $90.9(2)^\circ$, referenced to the $z$ coils, is compensated, but the uncertainty is included in the error budget.
No magnetic field is created along $x$, though a small, $\simeq \! 190 \mG$, uncompensated residual field along that direction contributes to the systematic uncertainty.
Stray magnetic fields in the other directions are compensated to $\lesssim \! 10 \mG$.
The uncertainty in the lattice beam polarization is conservatively estimated by means of the transmission ratio of a polarizing beam splitter, resulting in a $0.07^\circ$ contribution.
However, this value also needs to include the relative angle between the normal vector to the breadboard and the $z$-direction of the coils, which mainly depends on the coil manufacturing and assembly process.
We estimate $1 \mm$ uncertainty in the lateral displacement between the $z$ coil pair, resulting in a $0.92^\circ$ uncertainty. This is the dominating factor of the total magnetic angle systematic error: combining the various independent uncertainties using the root sum square of the individual contributions, yields a total systematic uncertainty of $0.93^\circ$.
\begin{figure*}[t!]
	\includegraphics{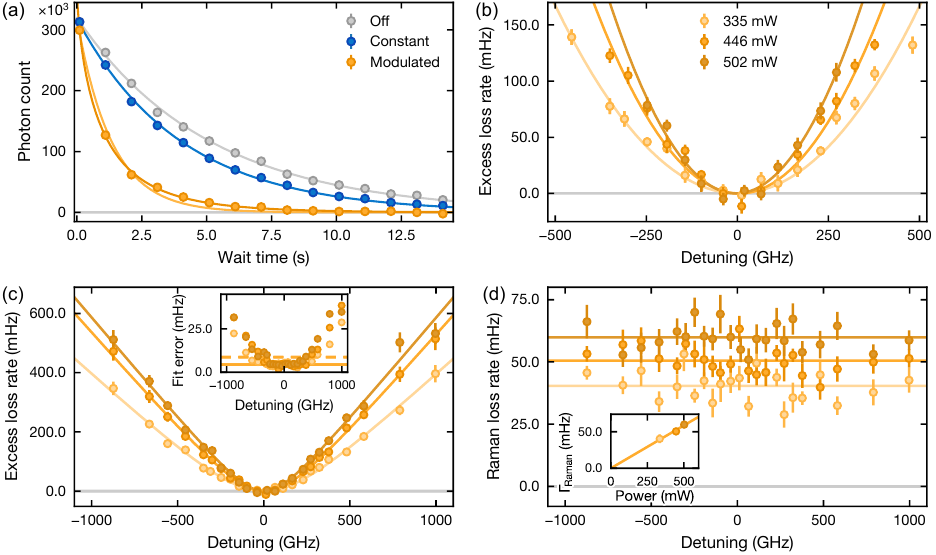}
	\caption{\label{fig:tuneout-systematics}
		\textbf{Assessing the systematic uncertainty from heating saturation effects and Raman losses}.
		$\left(\text{a}\right)$ At large detunings and for a strong modulation, the total evolution of atom loss can no longer be described by a single exponential fit (light orange solid line). Instead, a super-exponential fit captures the time dynamics well (dark orange solid line). The loss rates without (gray) and an unmodulated (blue) tune-out lattice are not affected; the two solid lines show a single exponential fit.
		$\left(\text{b}\right)$ Restricted data set based on the fit uncertainty as described in the text and displayed in $\left(\text{c}\right)$ as an inset, to limit saturation effects. The data is fitted with a quadratic fit function without offset (solid lines). The data at a tune-out power of $446 \mW$ is shown in the main text.
		$\left(\text{c}\right)$ Complete dataset, which shows a linear evolution for large detunings, necessitating fits with the empirical function \eqref{eq:fitfunction} to obtain bounds on the systematic error. Inset: The single exponential fit uncertainty propagated from the modulated and unmodulated lifetime fits remains approximately constant for small detunings and rapidly diverges once the saturation effect starts to become relevant. We define the acceptance threshold (dashed line) to be twice the mean fit uncertainty for small detunings (solid line).
		$\left(\text{d}\right)$ Upon comparing the lifetime in the presence of a constant tune-out beam and the one without, we do not observe any significant trends as a function of laser detuning.
        The solid lines represent the means of the Raman loss rates for each power level (color coded as in the other subfigures and the inset). Inset: The dependence of the mean Raman loss rate on the power is linear, in agreement with Fig.~\subfigref{Raman}{a}.}
\end{figure*}

\section{Systematic uncertainties in the tune-out measurement}
Since saturation effects can play a significant role in the determination of tune-out wavelengths by means of parametric heating schemes \cite{hoehn:2023}, we perform three lifetime measurements with varying modulation powers at each selected tune-out laser frequency.
This results in varying detuning ranges, at which the saturation effect starts to become relevant.
We determine this onset using the exponential fit uncertainty of the modulated lifetime curves, which starts to increase beyond the base value of $\sigma \simeq \! 4 \mHz$ due to deviations from a single exponential decay [Fig.~\subfigref{tuneout-systematics}{a}].
We attribute this to be the result of spatial inhomogeneities in the parametric heating due to the incommensurability of the lattices, combined with strongly varying tunneling rates depending on the harmonic oscillator state an atom occupies.
For a strong modulation, where the perturbative approach fails, this leads to rapid loss of initially relatively hot atoms independent of their initial position, while the coldest atoms can remain cold for comparably long times if they reside on lattice sites that do not observe strong effective amplitude modulation.
The resulting loss trend can be approximated by a super-exponential function $N_0 e^{-(\Gamma t)^\alpha}$, which indeed captures the data well.
For the tune-out value mentioned in the main text we select an upper bound of the fit uncertainty of $2 \sigma = 8.8 \mHz$, beyond which all datapoints are excluded.
Since we moreover select symmetrically spaced detunings around the tune-out wavelength, which are additionally chosen to be equidistant close to $f_\mathrm{to}$, we choose a symmetrically truncated dataset to prevent bias.
This leads to a dataset that contains 17 distinct wavelengths for the lowest optical power of $\simeq \! 335 \mW$, 15 for $446 \mW$, and 11 for $502 \mW$ [Fig.~\subfigref{tuneout-systematics}{b}].
For these subsets, we find the expected simple quadratic fit without offset to describe the data well, as confirmed with a $\chi^2$ test.
For an in-depth discussion of the perturbative derivation of this functional dependence we refer to Refs.~\cite{savard:1997,heinz:2020}.
To obtain an estimate of the systematic uncertainty due to undetected and uncorrected heating saturation, we additionally perform a fit to these subsets as well as to the complete datasets with the function
\begin{equation}\label{eq:fitfunction}
	\Gamma\!_\mathrm{exc}(\Delta) = A \left(\sqrt{B \Delta^2 + 1} - 1\right),
\end{equation}
where $A$ and $B$ are the fitting parameters, that describes a quadratic function for small detunings and becomes linear for large $\Delta$.
This empirical function takes the deviation from the perturbative quadratic scaling result due to the emergence of an additional loss timescale into account.
The extremal discrepancies from $f_\mathrm{to}$ amount to $+ 5.7 \GHz$ and $- 4.5 \GHz$ for the full and restricted dataset, respectively, and we conservatively assume these values to quantify the systematic uncertainty to the tune-out measurement.

We exclude other sources of potential systematic uncertainties, such as different mean intensity levels for the modulated and the constant case, by analyzing sample photodiode traces.
While we notice a slightly larger mean photodiode voltage upon modulation, this effect only amounts to a relative difference of $0.4(9)\%$, i.e., within the measurement uncertainty, and corresponds to an enhanced Raman loss rate of $0.2(5) \mHz$ for the dataset at $502 \mW$.
The inclusion of a corresponding offset into the fit function further does not affect the results noticeably and is therefore not applied for the values reported in the main text.
Moreover, the computed Raman loss rate only varies by $5\%$ over the whole detuning range of $1.9 \THz$, allowing us to treat it as approximately constant.
Likewise, the deviation of the calculated polarizability curve from a linear function due to its curvature amounts to less than $1\%$ and is therefore neglected.

\end{document}